# X-RAY STANDING WAVES AT THE TOTAL REFLECTION CONDITION: DIRECT METHOD AND COHERENCE EFFECTS

MICHAEL J. BEDZYK

*Departments of Materials Science & Engineering and Physics & Astronomy, Northwestern University, Evanston, IL  60208, USA*

*and Argonne National Laboratory, Argonne, IL  60439, USA*

Fresnel theory is used to derive the complex electric-fields above and below an X-ray reflecting interface that separates two materials with differing indices of refraction. The interference between the incident and reflected waves produces an X-ray standing wave (XSW) above the reflecting interface.  The XSW intensity modulation is strongly enhanced by the total external reflection (TR) condition, which occurs at incident angles less than the critical angle. At these small milliradian incident angles the XSW period ($\lambda/_{2\theta}$) becomes very large, which makes the TR-XSW an ideal probe for studying low-density structures that extend 1 to 1000 nm above the reflecting interface. Fourier inversion of the XSW induced modulation in the X-ray fluorescence (XRF) yield from a specific atomic distribution within the overlayer directly produces a model-independent 1-D atomic density profile. The modulation can also be used to analyze the degree of coherence in the incident X-ray beam.

## 1.  Introduction

The original (and most widely-used) method for generating an x-ray standing wave (XSW) has been to use dynamical diffraction from a perfect single crystal in a Bragg reflection geometry [1-3]. However, as with any standing wave phenomena, the minimum requirement is the superposition of two coherently-coupled plane waves. Therefore, one can imagine several alternative geometries for generating an XSW. This chapter will discuss the case of generating an XSW by total external reflection from an x-ray mirror surface [4].

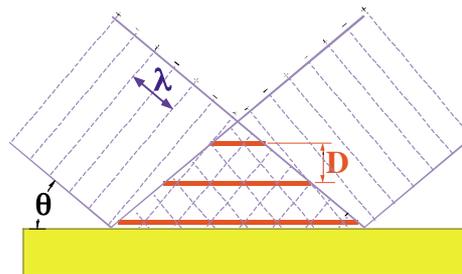

Figure 1. Illustration of XSW generated by interference between incident and specular reflected plane waves.

Referring to Fig. 1, the primary distinguishing feature for the total-reflection (TR) case is that the length of the XSW period above the mirror surface,

$$D = \frac{\lambda}{2\sin\theta} = \frac{2\pi}{Q} \quad , \tag{1}$$

is much longer, since TR occurs at very small incident angles, $\theta$. Also, the length of the XSW period, $D$, will continuously decrease as $\theta$ increases through the range of TR. This long-period XSW is ideally suited for measuring surfaces, interfaces, and supported nanostructures with structural features that range from 50 to 2000 Å. Examples include studies of Langmuir-Blodgett (LB) multi-layers [4-7], layer-by-layer self-assembly of metal-organic films [8, 9], the diffuse double-layer formation at the electrified water / solid interface [10, 11], biofilm ion adsorption [12], and metal nanoparticle dispersion in polymer films [13, 14].

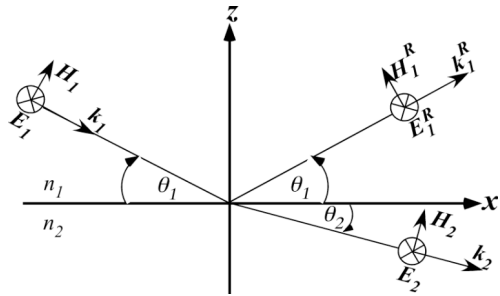

Figure 2. The σ-polarization case for the reflection and refraction of X-rays at a boundary separating two media with indices of refraction $n_1 > n_2$.

## 2. X-ray Transmission and Reflection at a Single Interface

Based on Maxwell's equations, an electromagnetic traveling plane-wave impinging on a boundary separating two different refractive media, splits into a reflected and transmitted (or refracted) plane-wave [15]. At X-ray frequencies the index of refraction,

$$n_j = 1 - \delta_j - i\beta_j \tag{2}$$

is less than unity and therefore (as illustrated in Fig. 2) the angle of refraction, $\theta_2$, is less than the incident angle, $\theta_1$ [16]. Parameters $\delta_j$ and $\beta_j$, which account respectively for refraction and absorption effects by the $j^{th}$ medium, can be expressed as:

$$\delta = -\frac{1}{2}\chi_0' = \frac{r_e \lambda^2}{2\pi} N_e' \quad , \tag{3}$$

$$\beta = -\frac{1}{2}\chi_0'' = \frac{\lambda \mu_0}{4\pi} \quad , \tag{4}$$

where $N_e'$ is the real part of the effective electron density.

The E-fields associated with the incident, reflected, and transmitted plane-waves are expressed respectively as:

$$\bar{\mathcal{E}}_1(\mathbf{r},t) = \mathbf{E}_1 \exp(-i(\mathbf{k}_1 \cdot \mathbf{r} - \omega t)) \qquad (5a)$$

$$\bar{\mathcal{E}}_1^R(\mathbf{r},t) = \mathbf{E}_1^R \exp(-i(\mathbf{k}_1^R \cdot \mathbf{r} - \omega t)) \qquad (5b)$$

$$\bar{\mathcal{E}}_2(\mathbf{r},t) = \mathbf{E}_2 \exp(-i(\mathbf{k}_2 \cdot \mathbf{r} - \omega t)) \qquad (5c)$$

At $z = 0$ the space and time variations of all three fields must be equivalent. This produces the "law of co-planarity", which requires the transmitted and reflected wave-vectors, $\mathbf{k_2}$ and $\mathbf{k_1}^R$, to be confined to the same plane as the incident wave-vector, $\mathbf{k_1}$ (the $xz$-plane in Fig. 2). The continuity of the tangential components of the three wave vectors at the boundary dictates the kinematical properties corresponding to the "law of reflection" $\theta_1^R = \theta_1$ and the "law of refraction" (Snell's Law) $n_2 \cos\theta_2 = n_1 \cos\theta_1$. Using these relationships the spatial components in Eq. (5) can be expressed as:

$$\mathbf{k}_1 \cdot \mathbf{r} = k_1(x\cos\theta_1 - z\sin\theta_1) \qquad (6a)$$

$$\mathbf{k}_1^R \cdot \mathbf{r} = k_1(x\cos\theta_1 + z\sin\theta_1) \qquad (6b)$$

$$\mathbf{k}_2 \cdot \mathbf{r} = \frac{n_2}{n_1} k_1 \left[ x \frac{n_1}{n_2} \cos\theta_1 - z\sqrt{1 - (\frac{n_1}{n_2}\cos\theta_1)^2} \right] \qquad (6c)$$

Total reflection occurs when the transmitted plane-wave $\bar{\mathcal{E}}_2(\mathbf{r},t)$ propagates strictly in the $x$-direction and is attenuated in the inward negative $z$-direction. From Eqs. (5c) and (6c), TR occurs when $\theta_1 < \theta_C$. For $n_1 = 1$ (e.g., vacuum or air) and $n_2 = 1 - \delta - i\beta$, the critical angle [16] is:

$$\theta_C = \sqrt{2\delta} \ . \qquad (7)$$

The scattering vector at the critical angle is

$$Q_c = 4\pi\sin\theta_c / \lambda \cong 4\pi\theta_c / \lambda = 4\sqrt{\pi \ r_e N_e'} \ . \qquad (8)$$

If dispersion corrections are ignored, $N_e' = N_e$ and $Q_c$ becomes a wavelength-independent property. In which case, $Q_c = 0.0315 \ Å^{-1}$ for Si and $Q_c = 0.0812 \ Å^{-1}$ for Au.

The continuity of the tangential components of the E-fields and magnetic-fields at the $z = 0$ boundary dictates the dynamical properties of the fields, corresponding to the Fresnel equations, which for the σ-polarization case and for small angles $\theta_1$ can be expressed as:

$$F^R_{1,2} = \frac{E^R_1}{E_1} = \left|\frac{E^R_1}{E_1}\right|e^{iv} = \frac{q - \sqrt{q^2 - 1 - i\,b}}{q + \sqrt{q^2 - 1 - i\,b}} \qquad (9)$$

$$F^T_{1,2} = \frac{E_2}{E_1} = \frac{2q}{q + \sqrt{q^2 - 1 - i\,b}}, \qquad (10)$$

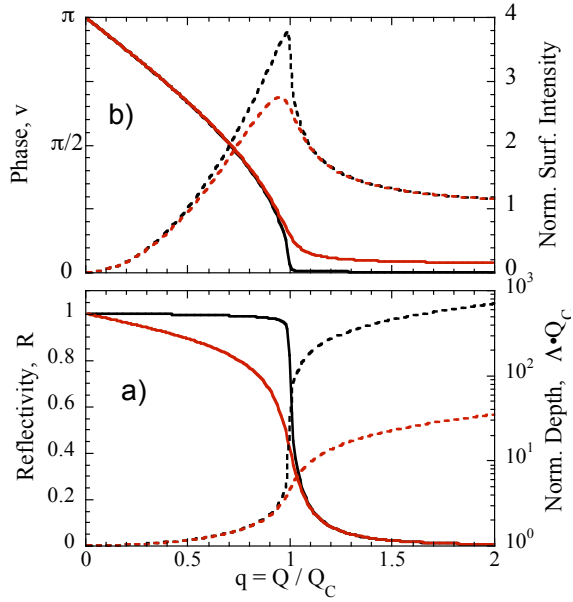

where the normalized angle $q = \theta_1/\theta_C = Q/Q_c$ and $b = \beta/\delta$ for the case of $n_1 = 1$ and $n_2 = 1 - \delta - i\beta$. At $\lambda = 0.71$ Å, $b = 0.005$ for Si and $b = 0.1$ for Au.

Figure 3 shows the $q$ dependence of the reflectivity, $R = \left|F^R_{1,2}\right|^2$, the normalized E-field intensity at the surface, $I_{z=0} = \left|F^T_{1,2}\right|^2$, the phase of the reflected plane-wave relative to the phase of the incident at $z = 0$, $v = Arg\left(F^R_{1,2}\right)$, and the penetration depth.

Figure 3. The Fresnel theory calculated, normalized-angular dependence of the **a)** reflectivity (solid lines) and normalized penetration depth (dashed lines) and **b)** phase (solid lines) and normalized surface E-field intensity (dashed lines). The black colored curves are for the weak absorption-case of b = 0.005 and the red curves are for b = 0.1.

$$\Lambda = 1/\mu_e = \left[ Q_c \ \text{Re} \left[ \sqrt{1 - q^2 + i\ b} \ \right] \right]^{-1} . \quad (11)$$

As can be seen TR occurs for *q < 1*, where the reflectivity approaches unity, the phase shifts by π radians, and E-field intensity below the surface forms an evanescent wave [17] with a penetration depth approaching $Q_c^{-1}$, which is 32 Å for Si and 12 Å for Au. For *q* increasing above unity, the reflectivity quickly reduces (approaching zero) and the transmitted wave propagates into the medium with a penetration depth quickly approaching the normal absorption process, where *Λ = sinθ/μ₀*.

An equivalent expression for the complex reflectivity amplitude of Eq. (9) can be derived from dynamical diffraction theory by solving for the symmetric $0^{th}$-order Bragg diffraction condition. i.e., set the structure factor $F_H = F_0$ in the expression for angle parameter *η*. This equivalence is simply due to the fact that TR is the $0^{th}$-order dynamical Bragg diffraction condition, where the d-spacing is infinite.

## 3. The E-Field Intensity

The total E-field in the vacuum (or air) above the mirror surface, where the incident and reflected plane waves are coherently coupled by $\mathbf{Q} = \mathbf{k_1^R} - \mathbf{k_1}$, is expressed as $\bar{\varepsilon}_T = \bar{\varepsilon}_1 + \bar{\varepsilon}_1^R$, and below the mirror surface, $\bar{\varepsilon}_T = \bar{\varepsilon}_2$. The E-field intensity, $I = |\bar{\varepsilon}_T|^2$, can then be expressed as:

$$I(\theta,z) = I_0 \begin{cases} 1 + R + 2\sqrt{R} \cos(v - Qz), & for \ z \geq 0 \\ \left|F_{1,2}^T\right|^2 \exp(-\mu_e|z|), & for \ z \leq 0 \end{cases}, \quad (12)$$

where $I_0 = |E_1|^2$ is the intensity of the incident plane wave and $\mu_e$ is the effective linear absorption coefficient defined in Eq. (11). As can be seen in Fig. 4 the E-field intensity under the TR condition exhibits a standing wave above the mirror surface with a period *D = 2π/Q* and an evanescent wave below the surface. The height coordinate in Fig. 4 is normalized to the critical period $D_c = 2\pi/Q_c$, which is 199 Å for Si and 77 Å for Au (if *Δf " = Δf ' = 0*).

As can be seen from Figs. 3 and 4, at *q = 0* there is a node in the E-field intensity at the mirror surface and the first antinode is at infinity. As *q* increases, that first antinode moves inward and reaches the mirror surface at *q = 1*. This inward movement of the first antinode, which is analogous to the Bragg diffraction case, is due to the π phase-shift depicted in Fig 3b. The other XSW antinodes follow the first antinode with a decreasing period of $D = 2\pi/Q$. For $q$ increasing above unity, the XSW phase is fixed, the period $D$ continues to contract, and the XSW amplitude drops off dramatically.

## 4. X-Ray Fluorescence Yield from an Atomic Layer within a Thin Film

The $q$ dependence for the normalized E-field intensity at $z = 0$ is shown in Fig. 3b. Figure 5 shows the Eq. (12) calculation for two additional heights above the surface. These three curves illustrate the basis for the TR-XSW technique as a positional probe, since (in the dipole approximation for the photo-effect) the XRF yield, $Y(q)$, from an atomic-layer at a discrete height z will follow such a curve. Note that in the TR range, $0 < q < 1$, the number of modulations in the E-field intensity is equivalent to $z/D_c + ½$. The extra ½ modulation is due to the π phase shift shown in Fig. 3b. Therefore, for an XRF marker atom layer within a low-density film on a high-density mirror, the atomic layer height can be quickly approximated by counting the number of modulations in the XRF yield that occur between the film critical angle and the mirror critical angle. Referring to Fig. 6, this effect can be seen in the experimental results and analysis for the case of a Zn atomic layer trapped at the topmost bilayer of a 1000 Å thick LB multilayer that was deposited on a gold mirror. There are 11½ Zn Kα XRF modulations as the incident angle is advanced over this range, indicating that the Zn layer is at a height of *11 critical periods* (or *900 Å*) above the gold surface. From the simultaneously collected reflectivity shown in Fig. 6, the critical angles for the LB film and Au mirror are at *2.15* and *7.52 milli-radians*, respectively. A more rigorously determined Zn atomic distribution profile, $\rho(z)$, is determined by a fit of the modeled XRF yield $Y(\theta) = \int \rho(z) I(\theta,z) dz$ to the data in Fig. 6, where the E-field intensity $I(\theta,z)$ within the refracting (and absorbing) film was calculated by an extension of Parratt's recursion formulation[18] described in the section entitled "XSW in Multilayers". This same model described in the inset was also used to generate a fit to the reflectivity data (Fig. 6), which is independently sensitive to the density and thickness of the film and the widths of the interfaces. The very sharp drop in the reflectivity at the film critical angle (*2.15 mrad*) is due to the excitation of the first mode of a resonant cavity that was observed to produce a 20-fold enhancement in the E-field intensity at the center of the film [19].

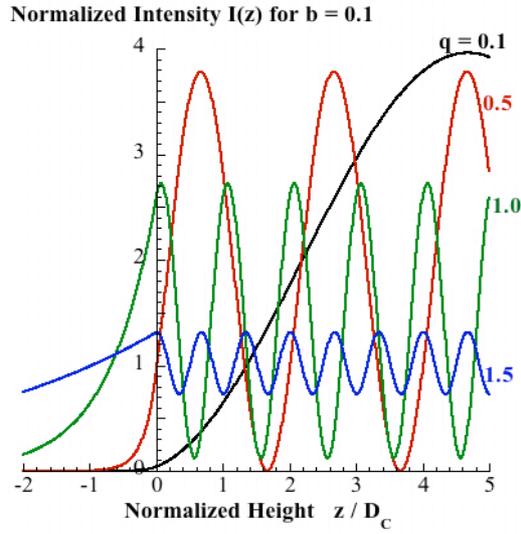

Figure 4. The normalized-height dependence of the normalized E-field intensity for different normalized incident angles $q$. An XSW exists above the mirror surface and an evanescent wave exists below the surface for $q < 1$. The calculation is for the case of $b = 0.1$ in Eq. (12).

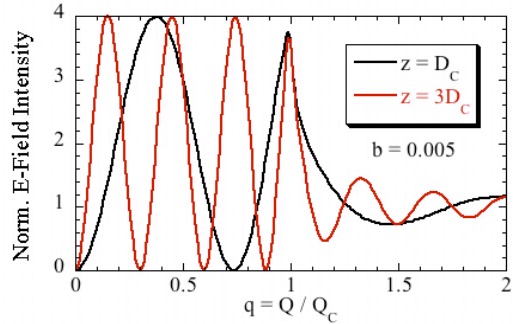

Figure 5. The normalized-angle dependence of the normalized E-field intensity for two different heights above the mirror surface.. The calculations are for the case of $b = 0.005$ in Eq. (12). The critical period $D_C = 199$ Å for Si and $77$ Å for Au.

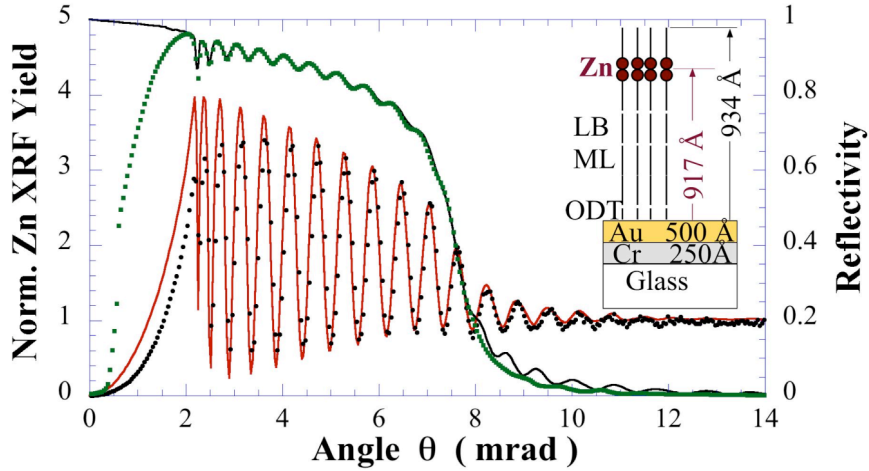

Figure 6. The experimental and theoretical reflectivity and Zn Kα XRF yield vs. incident angle at $\lambda = 1.215\ \text{Å}$ from a LB multilayer-coated gold mirror depicted in the inset. From the reflectivity fit the film thickness is measured to be $t_F = 934\ \text{Å}$ and the interface roughness $\sigma_I = 3\ \text{Å}$. From the XSW Zn Yield fit with a modeled Zn Gaussian distribution, $<z> = 917\ \text{Å}$ and $\sigma_{Zn} = 15\ \text{Å}$ (*FWHM = 35.3 Å*). The data deviation from theory for $\theta < 2\ mrad$ is due to x-ray foot-print geometrical effects. See Ref. [5] for details from a similar measurement on a similar sample.

## 5. Fourier Inversion for a Direct Determination of ρ(z)

Similar to the Bragg diffraction XSW case (See the chapter entitled "XSW Imaging".), the TR-XSW XRF yield is also directly linked to the Fourier transform of the atomic distribution. In this case, however, the Fourier transform is measured at low-$Q$, over a continuous range in $Q$ and only in the $Q_z$ direction. This section describes how the Fourier transform can be extracted from the TR-XSW data to produce a model-independent measure of the atomic distribution profile, $\rho(z)$.

To account for the refraction and absorption effects that will influence the observed reflectivity and XRF yields from a film coated mirror, the earlier described two-layer model (Fig. 2) needs to be replaced by a three-layer model (or double interface model) formed by vacuum (the $j = 1$ layer), a thin low-density film ($j = 2$), and a higher-density mirror ($j = 3$). The $j = 2$ film/mirror interface is at $z = 0$ and the $j = 1$ vacuum/film interface is at $z = t_F$. For the present case study, $\delta_1 = \beta_1 = 0$, $\delta_2 << \delta_3$, and $\beta_2 << \beta_3 << \delta_3$. For the Fig. 6 example of *1.215 Å* x-rays reflecting from a gold mirror coated with a LB multilayer, $\delta_2 = 2.31 \times 10^{-6}$, $\delta_3 = 2.83 \times 10^{-5}$, $\beta_2 = 1.90 \times 10^{-9}$, $\beta_3 = 1.9610^{-6}$, $\theta_{c,2} = 2.15$ *mrad*, $\theta_{c,3} = 7.52$ *mrad*. At $\theta_1 = \theta_{c,3}$ the XSW period is *80.8 Å* in the vacuum and *84.2 Å* inside the LB film.

TR exists at the interface above the $j^{th}$ layer when $\theta_1 < \theta_{c,j} = (2\delta_j)^{1/2}$. When $\theta_1 > \theta_{c,2}$, a refracted (or transmitted) traveling wave penetrates through the film and is reflected from the mirror surface. The total E-field intensity within the film is then described as:

$$I_2^T(q_2,Z) = I_2(q_2,Z) + I_2^R(q_2,Z) + 2\sqrt{I_2(q_2,Z)\ I_2^R(q_2,Z)}\ \cos(v_2(q_2) - 2\pi q_2 Z) \quad (13)$$

where $I_2$ and $I_2^R$ are the respective intensities of the incident (transmitted) and reflected plane waves that form an interference field within the film. The refraction-corrected normalized angle (or normalized scattering vector) within the film is defined as:

$$q_2 = Q_2/Q_{c,2} = \left(\theta_1^2 - \theta_{c,2}^2\right)^{1/2}\left(\theta_{c,3}^2 - \theta_{c,2}^2\right)^{-1/2}. \quad (14)$$

Z is the normalized height above the mirror surface in units of the refraction corrected critical period. Namely, $Z = z/D_{c,2}$, where $D_{c,2} = \lambda_1/(2\theta_{c,3})/(1 - \frac{1}{2}\delta_2/\delta_3)$. The use of generalized coordinates $q_2$ and $Z$ makes the description independent of wavelength and index of refraction. The phase of the reflected plane wave relative to the incident at $z = Z = 0$ is expressed as $v_2$.

Based on the dipole approximation for the photoelectric effect, the fluorescence yield from a normalized atomic distribution $\rho(Z)$ within the film is:

$$Y(q_2) = \int_0^{t'_F} I_2^T(q_2, Z)\, \rho(Z)\, dZ, \qquad (15)$$

where $t'_F = t_F/D_{c,2}$ is the normalized film thickness.

**Weakly Absorbing Film Approximation:** If the attenuation depth of $I_2$ and $I_2^R$ within the film is large in comparison to the spread $<Z^2>^{1/2}$ of $\rho(Z)$, then Eq. (15) can be simplified, so that the yield for a normalized incident intensity and normalized distribution is expressed as:

$$Y(q_2) = I_2(q_2, <Z>) + I_2^R(q_2, <Z>) + 2\sqrt{I_2(q_2, <Z>)I_2^R(q_2, <Z>)}\, y(q_2), \quad (16)$$

where the modulation in the yield due to the interference fringe field is:

$$y(q_2) = \int_0^{t'_F} \rho(Z)\, \cos(v_2(q_2) - 2\pi q_2 Z)\, dZ. \qquad (17)$$

Since $I_2(q_2, <Z>)$ and $I_2^R(q_2, <Z>)$ can be calculated from Parratt's recursion formulation, this reduced yield, $y(q_2)$, can be extracted from the measured yield $Y(q)$. Figure 7 shows this for the yield data shown in Fig. 6. The inverse Fourier transform of this reduced yield can be directly used to generate the fluorescence selected atom distribution $\rho(z)$ to within a resolving limit defined by the range of $Q$ over which the visibility of the interference fringes is significant.

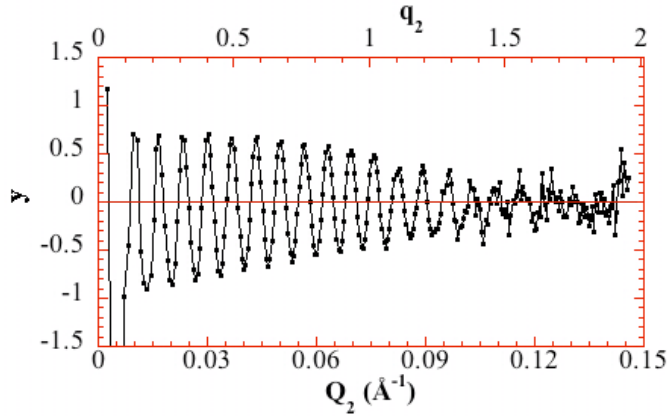

Figure 7. The reduced yield that was extracted from the $Zn\,K\alpha$ yield data in Fig. 6. See Eqs. (16)-(17). The line connecting data points is drawn to guide the eye. The oscillation period and envelop width are inversely related to the mean height and intrinsic width, respectively, of the Zn distribution profile.

**Linear Phase Approximation:** If the phase $v_2$ were zero in Eq. (17), $y(q_2)$ would simply be the real part of the Fourier transform of $\rho(Z)$. Based on Fresnel theory for the case of no absorption *($\beta = 0$)* the phase is:

$$v_2(q_2) = \begin{cases} \cos^{-1}(2q_2^2 - 1) & ,\text{for } 0 \leq q_2 < 1 \\ 0 & ,\quad\quad\text{for } q_2 \geq 1 \end{cases}. \quad (18)$$

As can be seen in Fig. 3b, $v$ can be reasonably approximated by a linear function in the TR region as: $v(q) \sim (1 - q)\pi$ , for $0 \leq q < 1$. Introducing this approximation into Eq. (17), simplifies the expression for the reduced yield to:

$$y(q_2) = \begin{cases} -\int_0^1 \rho(Z)\cos[2\pi q_2(Z + 1/2)]dZ & ,\text{ for } 0 \leq q_2 < 1 \\ \int_0^1 \rho(Z)\cos[2\pi q_2 Z]dZ & ,\quad\quad\text{ for } q_2 \geq 1 \end{cases}. \quad (19)$$

The atomic density profile can then be directly generated from the TR-XSW data as:

$$\rho(Z) = \sum_{q_2 > 0} s(q_2) y(q_2) \cos[2\pi q_2(Z + \delta(q_2))] \Delta q_2 , \quad (20)$$

where $s = -1$ and $\delta = \frac{1}{2}$ for $0 < q_2 < 1$ and $s = 1$ and $\delta = 0$ for $q_2 > 1$. In Fig. 8 this is illustrated for the data in Figs. 6 and 7. The resolution for this model-independent Fourier inversion of this data is $\pi/Q_{2,max} = 25$ Å. The precision for the height and width of a Gaussian model fit to this type of data is typically *±2 to ±5 Å* [4, 5].

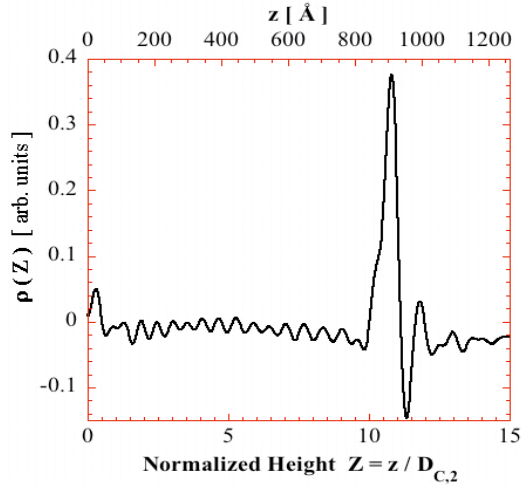

Figure 8. The *Zn* distribution profile directly generated by the Fourier inversion of the reduced *Zn K$\alpha$* XRF yield data from Fig. 7. The summation in Eq. (20) was for data in the range *$0.09 < q_2 < 1.91$*. The peak at *$Z = 10.8$ or $z = 909$ Å* has a *FWHM = 43 Å*. This corresponds to the convolution of the intrinsic width *(35 Å)* with the resolution width *(25 Å)*. The truncation-error oscillations have a period of *50 Å* corresponding to an effective $Q_2$ range of *$0.13$ Å$^{-1}$*.

## 6. The Effect of Coherence on X-ray Interference Fringe Visibility

If the spatial and temporal coherence properties of the incident photon beam are well known, the TR-XSW observation described above can be used to determine the spatial distribution of the fluorescent atom species within the film. Conversely, if the spatial distribution of the fluorescent atomic species is known, the observation of the x-ray interference fringes can be used to characterize the longitudinal and transverse coherence lengths of the incident photon beam. This is demonstrated in Fig. 9, where three separate TR-XSW measurements taken with three different longitudinal coherence lengths *($L_L = \lambda^2/\Delta\lambda$)* of the same LB multilayer structure described in the inset of Fig. 6. The fringe visibility, as observed by the *Zn K$\alpha$* fluorescence, is not affected by a reduction in $L_L$ until the optical path-length difference (in units of $\lambda$) between the two beams at the *Zn* height (expressed as $Q_2<z>/2\pi$) approaches the value of the monochromaticity, *$\lambda/\Delta\lambda$*. Referring to Fig. 10, the optical path-length difference is *$n(BC - AC) = n(2z\sin\theta)$*. In Fig. 9 the top curve (identical to Fig. 5) corresponds to a nearly ideal plane wave condition produced by using a *Si(111)* monochromator. The lower two curves correspond to much wider band-pass incident beams that were prepared by Bragg diffraction from two different multilayer monochromators (*Si/Mo* and *C/Rh*).

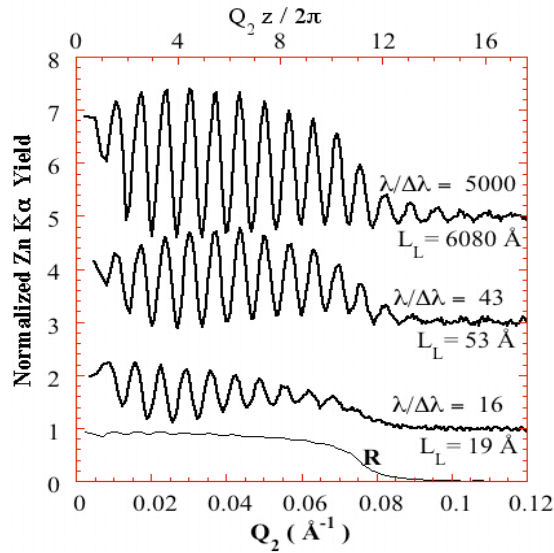

Figure 9. Experimental demonstration of TR-XSW sensitivity to longitudinal coherence *($L_L$)*. The bottom and top curves are the $Q_2$ dependence of the reflectivity *(R)* and *Zn* XRF yield for the data from Fig. 6 that was taken with a *Si(111)* monochromator. The mean *Zn* height is at *z = 917 Å*. The lower two XRF yield curves are from the same sample, but taken with *Si/Mo* and *C/Rh* multilayer monochromators with reduced monochromaticity *($\lambda/\Delta\lambda$)* and therefore reduced longitudinal coherence. The top two curves are vertically offset by 2 and 4 units, respectively.

A reduction in the interference fringe visibility due to a limited transverse coherence should not occur if the transverse coherence length $L_T >> z$. Therefore, in consideration of typical longitudinal and transverse coherence lengths at third generation SR undulator beamlines, the TR-XSW method that uses 1 Å wavelength x-rays should be extendable as a probe to a length-scale of 1 µm above the mirror surface.

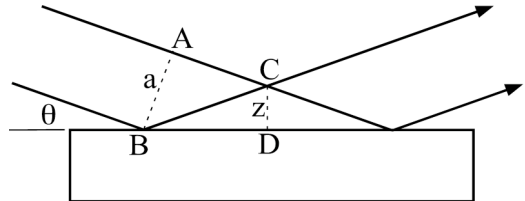

Figure 10. Schematic ray diagram used for illustrating coherence effects between the incident and specular reflected x-ray beams at height $z$ above the mirror surface.

It is worth noting that the reduced yield $y(q)$, as defined in Eqs. (16) and (17) and shown measured and Fourier inverted in Figs. 6 and 7, is a measurement of the real part of the degree of coherence $Re[\gamma_{12}]$ [15].

*Acknowledgements:*

Colleagues who inspired and assisted in this work include: Donald Bilderback, Boris Batterman, Martin Caffrey, Hector Abruna, Mark Bommarito, Jin Wang, Thomas Penner, Jay Schildkraut, Paul Fenter and Gordon Knapp. The data in this chapter was collected at the D-Line station of the Cornell High Energy Synchrotron Source (CHESS), which is supported by the US National Science Foundation. The work was also partially supported by the US Department of Energy.